\documentclass[letterpaper,twocolumn,conference]{IEEEtran}
\usepackage[T1]{fontenc}
\usepackage{bm}
\usepackage{amsmath}
\usepackage{amssymb}
\usepackage{graphicx}
\usepackage[unicode=true,
 bookmarks=true,bookmarksnumbered=true,bookmarksopen=true,bookmarksopenlevel=1,
 breaklinks=false,pdfborder={0 0 0},pdfborderstyle={},backref=false,colorlinks=false]
 {hyperref}
\hypersetup{pdftitle={Your Title},
 pdfauthor={Your Name},
 pdfpagelayout=OneColumn, pdfnewwindow=true, pdfstartview=XYZ, plainpages=false}

\makeatletter

\pdfpageheight\paperheight
\pdfpagewidth\paperwidth

\providecommand{\tabularnewline}{\\}

\usepackage{cite}
\usepackage[caption=false,font=footnotesize]{subfig}

\@ifundefined{showcaptionsetup}{}{%
 \PassOptionsToPackage{caption=false}{subfig}}
\usepackage{subfig}
\makeatother

\begin{document}

\title{Vehicle-to-Vehicle Communications with Urban Intersection Path Loss
Models}

\author{Mouhamed~Abdulla, Erik~Steinmetz, and Henk~Wymeersch\\
Department of Signals and Systems\\
Chalmers University of Technology, Sweden\\
Email: \{mouhamed,estein,henkw\}@chalmers.se}
\maketitle
\begin{abstract}
Vehicle-to-vehicle (V2V) communication can improve road safety and
traffic efficiency, particularly around critical areas such as intersections.
We analytically derive V2V success probability near an urban intersection,
based on empirically supported line-of-sight (LOS), weak-line-of-sight
(WLOS), and non-line-of-sight (NLOS) channel models. The analysis
can serve as a preliminary design tool for performance assessment
over different system parameters and target performance requirements.
\end{abstract}

\section{Introduction\label{S1: Introduction}}

According to the UN\textquoteright s World Health Organization, around
1.25 million road-traffic deaths occur every year \cite{WHORoadSafety}.
Moreover, it is worth remarking that a significant fraction of these
fatalities occur at intersections \cite{nhtsa2}, due to careless
driving, speeding, driving under the influence, etc. On the technological
side, next-generation wireless systems have given a lot of attention
to the paradigm of vehicle-to-vehicle (V2V) communications, particularly
for the purpose of road safety and traffic efficiency. Indeed, support
for V2V services is already part of LTE Release 14, and this momentum
will further continue on as we gradually migrate to future networks
such as 5G. 

For road-safety purposes, packet reliability is a key performance
metric in the 5G ecosystem \cite{5GPPPVision}. As a means to evaluate
this performance metric at the physical (PHY) layer, it is important
to develop analytical expressions in order to identify the contribution
of the relevant parameters during the design of V2V communication
systems and to gain fundamental insights. Stochastic geometry is well-suited
to develop such expressions for vehicular communication \cite{Blaszczyszyn2009,Blaszczyszyn2012,Jeong2013,Tong,Erik2015_Globecom}.
Intersections were explicitly considered in \cite{Erik2015_Globecom},
though only for suburban and rural scenarios. For the analytical expressions
to have practical relevance, they must build on validated empirically
supported propagation measurements \cite{Tufvesson2011IEEEMag,Tufvesson2011TVT}.
Since urban intersections have particular propagation characteristics
\cite{BMW2011_VirtualRelay,Abbas2013_Sweden}, it is meaningful to
perform a dedicated analysis, complementing \cite{Erik2015_Globecom}.
\begin{figure}
\centering{}\includegraphics[width=1\columnwidth]{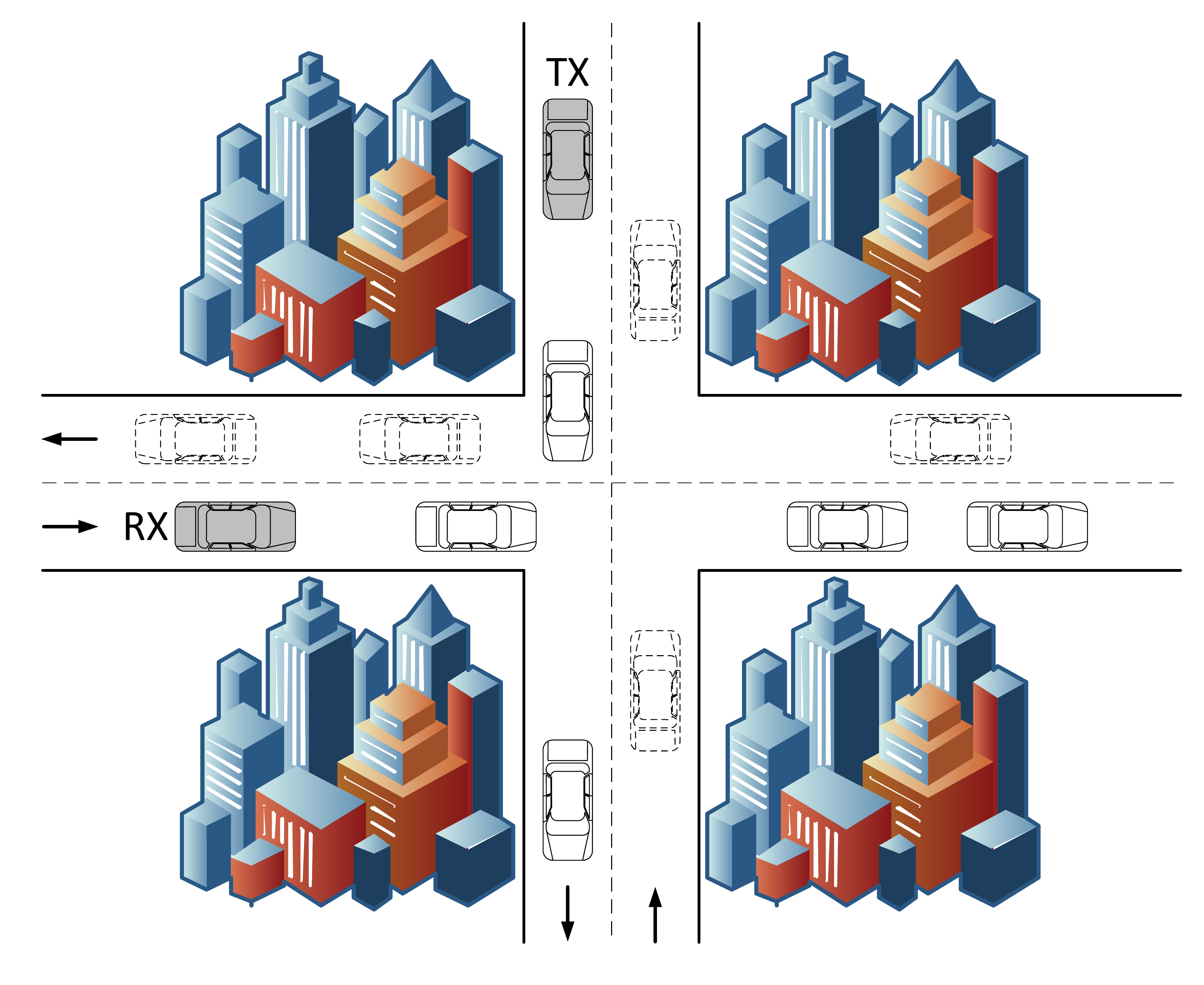}\caption{Characterization of the V2V intersection problem, whereby the transmitter
(TX) sends a data packet to a receiver (RX), in the presence of interfering
transmitters, over LOS, WLOS, and NLOS propagation environments.\label{Fig1: IntersectionPicture(visio)}}
\end{figure}

In this paper, we focus on the reliability of V2V communications around
urban intersections under line-of-sight (LOS), weak-line-of-sight
(WLOS), and non-line-of-sight (NLOS) scenarios, based on empirically
supported channel models. Our analysis is generic, considering a large
number of design parameters and system variables, and allows for closed-form
expressions for finite interference regions. We also provide design
guidelines in order to meet a target performance requirement. 

\section{System Model\label{S2: System Model}}

\subsection{Network Model}

We consider an intersection scenario, as depicted in Fig.~\ref{Fig1: IntersectionPicture(visio)},
comprising a transmitter (TX), which can be located anywhere on the
horizontal or vertical road, and a receiver (RX), which, without loss
of generality, is confined to the horizontal road. Hence, $\mathbf{x}_{\text{\ensuremath{\mathrm{tx}}}}=[x_{\mathrm{tx}},y_{\mathrm{tx}}]^{\mathrm{T}}$
and $\mathbf{x}_{\mathrm{rx}}=[x_{\mathrm{rx}},0]^{\mathrm{T}}$,
$x_{\mathrm{tx}},x_{\mathrm{rx}},y_{\mathrm{tx}}\in\mathbb{R}$, such
that $x_{\mathrm{tx}}y_{\mathrm{tx}}=0$. Other vehicles are randomly
positioned on both horizontal and vertical roads and follow a homogeneous
Poisson point process (H-PPP) over bounded sets $B_{\mathrm{x}}=\left\{ x\in\mathbb{R}\bigl|\left|x\right|\leq R_{\mathrm{x}}\right\} $
and $B_{\mathrm{y}}=\left\{ y\in\mathbb{R}\bigl|\left|y\right|\leq R_{\mathrm{y}}\right\} $,
with vehicular traffic intensities given respectively by $\lambda_{\mathrm{x}}$
and $\lambda_{\mathrm{y}}$. Interfering vehicles follow an Aloha
MAC protocol and can transmit independently with a probability $p_{\mathrm{I}}\in[0,1]$.
Hence, the interfering vehicles form thinned H-PPPs, denoted by $\Phi_{\mathrm{x}}\sim\textrm{PPP}\left(p_{\mathrm{I}}\lambda_{\mathrm{x}},B_{\mathrm{x}}\right)$
and $\Phi_{\mathrm{y}}\sim\textrm{PPP}\left(p_{\mathrm{I}}\lambda_{\mathrm{y}},B_{\mathrm{y}}\right)$.
 All vehicles, including TX, broadcast with the same power level
$P_{\circ}$. The receiver signal-to-interference-plus-noise-ratio
(SINR) threshold for reliable packet detection is set to $\beta$,
in the presence of additive white Gaussian noise (AWGN) with power
$N_{\circ}$. The SINR depends on the propagation channel, described
next. 

\subsection{Channel Model for Urban Intersection\label{S: CHANNEL MODEL}}

The received power observed at the RX from an active transmitter at
location $\mathbf{x}$ is modeled by $P_{\mathrm{rx}}(\mathbf{x},\mathbf{x}_{\mathrm{rx}})=P_{\circ}L_{\mathrm{ch}}(\mathbf{x},\mathbf{x}_{\mathrm{rx}})$,
which depends on transmit power $P_{\circ}$ and channel losses $L_{\mathrm{ch}}(\mathbf{x},\mathbf{x}_{\mathrm{rx}})$.
The channel losses consist of three components: deterministic path
loss $\ell(\mathbf{x},\mathbf{x}_{\mathrm{rx}})$ that captures the
propagation losses, random shadow fading $L_{\mathrm{s}}(\mathbf{x},\mathbf{x}_{\mathrm{rx}})$
that captures effects of obstacles, and random small-scale fading
$L_{\mathrm{f}}(\mathbf{x})$ that captures non-coherent addition
of signal components. For the purpose of tractability, we implicitly
consider shadow fading to be inherent within the H-PPP, and thus consider
$L_{\mathrm{ch}}(\mathbf{x},\mathbf{x}_{\mathrm{rx}})\simeq\ell(\mathbf{x},\mathbf{x}_{\mathrm{rx}})\,L_{\mathrm{f}}(\mathbf{x})$
\cite{Blaszczyszyn2013}. We model $L_{\mathrm{f}}(\mathbf{x})\sim\textrm{Exp}\left(1\right)$,
independent with respect to $\mathbf{x}$. In terms of the path loss,
we rely on measurements of V2V communication at 5.9 GHz for urban
intersections, which led to the so-called VirtualSource11p model \cite{BMW2011_VirtualRelay,Abbas2013_Sweden},
which serves as inspiration for our simplified model. For $\mathbf{x}$
on the same road as the RX (i.e., $\mathbf{x}=[x,0]^{\mathrm{T}}$),
our simplified model is
\begin{equation}
\ell\left(\mathbf{x},\mathbf{x}_{\mathrm{rx}}\right)=A_{\circ}\left\Vert \mathbf{x}_{\mathrm{rx}}-\mathbf{x}\right\Vert ^{-\alpha}=A_{\circ}|x_{\mathrm{rx}}-x|^{-\alpha},\label{eq:LOSmodel}
\end{equation}
which is a standard LOS Euclidean path loss. For $\mathbf{x}$ on
the orthogonal road (i.e., $\mathbf{x}=[0,y]^{\mathrm{T}}$), the
model is 
\begin{equation}
\ell\left(\mathbf{x},\mathbf{x}_{\mathrm{rx}}\right)=\begin{cases}
A_{\circ}^{\prime}\left(\left\Vert \mathbf{x}\right\Vert \Vert\mathbf{x}_{\mathrm{rx}}\Vert\right)^{-\alpha} & \min(|y|,|x_{\mathrm{rx}}|)>\triangle\\
A_{\circ}\left(\left\Vert \mathbf{x}\right\Vert +\Vert\mathbf{x}_{\mathrm{rx}}\Vert\right)^{-\alpha} & \min(|y|,|x_{\mathrm{rx}}|)\le\triangle,
\end{cases}\label{eq:WLOSNLOS}
\end{equation}
where the first case is relevant for NLOS communication, while the
second case should be used when either TX/interferer or RX are close
to the intersection, i.e., WLOS. In these expressions, $\Vert\cdot\Vert$
is the $l_{2}$-norm, $\alpha\!>\!1$ is the path loss exponent; $A_{\circ}$
and $A_{\circ}^{\prime}$ are suitable\footnote{$A_{\circ}$ can be estimated via the free-space path loss model operating
at frequency $f_{\circ}$, reference distance $d_{\circ}$, and generic
path loss exponent $\alpha$. Generally, $A_{\circ}^{\prime}\!<\!A_{\circ}\left(\triangle/2\right)^{\alpha}$
so that NLOS is more severe than WLOS and LOS propagation. } path loss coefficients, and $\triangle$ is the break-point distance,
typically on the order of the lane size (roughly 10 \textendash{}
15 m). We will only consider the case where the region of H-PPP interferers
is greater than the path loss break-point distance, i.e., $\min\left(R_{\mathrm{y}},R_{\mathrm{x}}\right)\geq\triangle$.

\subsubsection*{Remark}

The model in (\ref{eq:LOSmodel})\textendash (\ref{eq:WLOSNLOS})
exhibits discontinuities. A mixture (a linear weighting) of these
models can be used to avoid these discontinuities, though this is
not considered in this paper. 

\subsection{Problem Statement}

Our goal will be to determine the success probability $\mathcal{P}_{\mathrm{c}}\left(\beta,\mathbf{x}_{\text{\ensuremath{\mathrm{tx}}}},\mathbf{x}_{\mathrm{rx}}\right)=\Pr\left(\mathsf{SINR}\geq\beta\right)$,
i.e., the probability that the SINR is above the threshold $\beta$,
where 
\begin{align}
\mathsf{SINR} & =\frac{L_{\mathrm{f}}(\mathbf{x}_{\mathrm{tx}})\thinspace\ell\left(\mathbf{x}_{\text{\ensuremath{\mathrm{tx}}}},\mathbf{x}_{\mathrm{rx}}\right)}{{\displaystyle \sum_{\mathbf{x}\in\Phi_{\mathrm{x}}\cup\Phi_{\mathrm{y}}}L_{\mathrm{f}}\left(\mathbf{x}\right)\ell\left(\mathbf{x},\mathbf{x}_{\mathrm{rx}}\right)+\gamma_{\circ}}},\label{EQ2: SINR General Formula}
\end{align}
in which $\gamma_{\circ}$$=$$N_{\circ}/P_{\circ}$. We will abbreviate
$L_{\mathrm{f}}\left(\mathbf{x}_{\mathrm{tx}}\right)$ by $L_{\mathrm{f}}$
and we introduce $I=\sum_{\mathbf{x}\in\Phi_{\mathrm{x}}\cup\Phi_{\mathrm{y}}}L_{\mathrm{f}}\left(\mathbf{x}\right)\ell\left(\mathbf{x},\mathbf{x}_{\mathrm{rx}}\right)$.
We should remark that the performance results are solely based at
the PHY layer with basic point-to-point communications. There are
more advanced techniques that could further improve the performance
rate, such as: (i) spatial diversity, (ii) smart resource allocation,
(iii) low latency HARQ retransmission, (iv) high performance MAC protocols.

\section{Generalized Success Probability\label{S3: Generalized Success Probability}}

The success probability comprises several sources of randomness: interference
$I$ and the fading of the useful link $L_{\mathrm{f}}$. Hence,
\begin{alignat}{1}
 & \mathcal{P}_{\mathrm{c}}\left(\beta,\mathbf{x}_{\text{\ensuremath{\mathrm{tx}}}},\mathbf{x}_{\mathrm{rx}}\right)\nonumber \\
 & =\mathbb{E}_{I}\left\{ \Pr\left(L_{\mathrm{f}}\geq\beta\left(I+\gamma_{\circ}\right)/\ell\left(\mathbf{x}_{\text{\ensuremath{\mathrm{tx}}}},\mathbf{x}_{\mathrm{rx}}\right)\right)\right\} \nonumber \\
 & =\mathbb{E}_{I}\left\{ \exp\left(-\beta\left(I+\gamma_{\circ}\right)/\ell\left(\mathbf{x}_{\text{\ensuremath{\mathrm{tx}}}},\mathbf{x}_{\mathrm{rx}}\right)\right)\right\} ,\label{EQ3: Average Success}
\end{alignat}
where we have used the exponential distribution of the small-scale
fading. With $\beta^{\prime}=\beta/\ell\left(\mathbf{x}_{\text{\ensuremath{\mathrm{tx}}}},\mathbf{x}_{\mathrm{rx}}\right)$,
we obtain 
\begin{align}
 & \mathcal{P}_{\mathrm{c}}\left(\beta,\mathbf{x}_{\text{\ensuremath{\mathrm{tx}}}},\mathbf{x}_{\mathrm{rx}}\right)\nonumber \\
 & =\exp\bigl(-\beta^{\prime}\gamma_{\circ}\bigr)\mathbb{E}_{I}\negthinspace\left\{ \exp\bigl(-\beta^{\prime}I\bigr)\right\} .\label{EQ4: Success Probability}
\end{align}
We introduce $\mathcal{P}_{\mathrm{noint}}$$=$$\exp\bigl(-\beta^{\prime}\gamma_{\circ}\bigr)$,
which is the success probability in the absence of interference and
$\mathbb{E}_{I}\left\{ \exp\bigl(-\beta^{\prime}I\bigr)\right\} $
is the reduction of the success probability due to interference. Since
the interferers and their fading realization on the horizontal and
vertical roads are independent, we find that 
\begin{align}
 & \mathbb{E}_{I}\left\{ \exp\bigl(-\beta^{\prime}I\bigr)\right\} \label{eq:Interferencebreakdown}\\
 & =\mathbb{E}_{\Phi_{\mathrm{x}},L_{\mathrm{f}}}\left\{ \exp\bigl(-\beta^{\prime}\sum_{\mathbf{x}\in\Phi_{\mathrm{x}}}L_{\mathrm{f}}\left(\mathbf{x}\right)\ell\left(\mathbf{x},\mathbf{x}_{\mathrm{rx}}\right)\bigr)\right\} \nonumber \\
 & \times\mathbb{E}_{\Phi_{\mathrm{y}},L_{\mathrm{f}}}\left\{ \exp\bigl(-\beta^{\prime}\sum_{\mathbf{x}\in\Phi_{\mathrm{y}}}L_{\mathrm{f}}\left(\mathbf{x}\right)\ell\left(\mathbf{x},\mathbf{x}_{\mathrm{rx}}\right)\bigr)\right\} .\nonumber 
\end{align}
The two factors in (\ref{eq:Interferencebreakdown}), say $\mathcal{P}_{\mathrm{x}}$
and $\mathcal{P}_{\mathrm{y}}$, can be evaluated as 
\begin{align}
\mathcal{P}_{\mathrm{x}} & =\mathbb{E}_{\Phi_{\mathrm{x}}}\left\{ \mathbb{E}_{L_{\mathrm{f}}|\Phi_{\mathrm{x}}}\left\{ \prod_{\mathbf{x}\in\Phi_{\mathrm{x}}}\exp\bigl(-\beta^{\prime}L_{\mathrm{f}}\left(\mathbf{x}\right)\ell\left(\mathbf{x},\mathbf{x}_{\mathrm{rx}}\right)\bigr)\right\} \right\} \\
 & \stackrel{(a)}{=}\mathbb{E}_{\Phi_{\mathrm{x}}}\left\{ \prod_{\mathbf{x}\in\Phi_{\mathrm{x}}}\mathbb{E}_{L_{\mathrm{f}}}\left\{ \exp\bigl(-\beta^{\prime}L_{\mathrm{f}}\left(\mathbf{x}\right)\ell\left(\mathbf{x},\mathbf{x}_{\mathrm{rx}}\right)\bigr)\right\} \right\} \\
 & \stackrel{(b)}{=}\mathbb{E}_{\Phi_{\mathrm{x}}}\left\{ \prod_{\mathbf{x}\in\Phi_{\mathrm{x}}}\mathcal{L}\Bigl(\beta^{\prime}\ell\left(\mathbf{x},\mathbf{x}_{\mathrm{rx}}\right)\right\} ,
\end{align}
where transition (a) is due to the i.i.d.~nature of the small-scale
fading and the independence of the fading on the H-PPP. Transition
(b) includes the Laplace transform of the fading. For Rayleigh fading,
$\mathcal{L}\left(s\right)=1/\left(1+s\right)$.  After considering
the probability generating functional for an H-PPP \cite[p.86]{HaenggiBook2013},
we obtain
\begin{equation}
P_{\mathrm{x}}=\exp\Bigl(-\int\limits _{-R_{\mathrm{x}}}^{+R_{\mathrm{x}}}\frac{p_{\mathrm{I}}\lambda_{\mathrm{x}}}{1+1/(\beta^{\prime}\ell([x,0]^{\mathrm{T}},\mathbf{x}_{\mathrm{rx}}))}\mathrm{d}x\Bigr).\label{PXTHM}
\end{equation}
Similarly, $P_{\mathrm{y}}$ is obtained as 
\begin{equation}
P_{\mathrm{y}}=\exp\Bigl(-\int\limits _{-R_{\mathrm{y}}}^{+R_{\mathrm{y}}}\frac{p_{\mathrm{I}}\lambda_{\mathrm{y}}}{1+1/(\beta^{\prime}\ell([0,y]^{\mathrm{T}},\mathbf{x}_{\mathrm{rx}}))}\mathrm{d}y\Bigr).\label{PYTHM}
\end{equation}

\subsection{Contribution for Horizontal Road \textendash{} $\mathcal{P}_{\mathrm{x}}$}

To derive $\mathcal{P}_{\mathrm{x}}$, we substitute the channel model
of (\ref{eq:LOSmodel}) into (\ref{PXTHM}), and since $\mathbf{x}_{\mathrm{rx}}=\left[x_{\mathrm{rx}},0\right]^{\mathrm{T}}$
and $\mathbf{x}=\left[x,0\right]^{\mathrm{T}}$, the integration reduces
to:
\begin{align}
 & \mathcal{P}_{\mathrm{x}}=\exp\biggl(-\int\limits _{-R_{\mathrm{x}}}^{+R_{\mathrm{x}}}\frac{p_{\mathrm{I}}\lambda_{\mathrm{x}}}{1+\bigl(\left|x_{\mathrm{rx}}-x\right|/\zeta\bigr)^{\alpha}}\mathrm{d}x\biggr)\label{EQ8: Px}
\end{align}
such that $\zeta=\left(A_{\circ}\beta^{\prime}\right)^{1/\alpha}=\bigl(A_{\circ}\beta/\ell\left(\mathbf{x}_{\mathrm{tx}},\mathbf{x}_{\mathrm{rx}}\right)\bigr)^{1/\alpha}$.
We show in Appendix \ref{sec:Expression-forPx} that $\mathcal{P}_{\mathrm{x}}=\exp\bigl(-p_{\mathrm{I}}\lambda_{\mathrm{x}}\zeta\mathcal{X}(R_{\mathrm{x}})\bigr)$,
where

\begin{align}
\mathcal{X}\left(R_{\mathrm{x}}\right) & =g_{\circ}\Bigl(\alpha,\frac{\bigl(R_{\mathrm{x}}+\left\Vert \mathbf{x}_{\mathrm{rx}}\right\Vert \bigr)}{\zeta}\Bigr)\nonumber \\
 & +g_{\circ}\Bigl(\alpha,\frac{\bigl(R_{\mathrm{x}}-\left\Vert \mathbf{x}_{\mathrm{rx}}\right\Vert \bigr)}{\zeta}\Bigr)\bm{1}{}_{\left\Vert \mathbf{x}_{\mathrm{rx}}\right\Vert \leq R_{\mathrm{x}}}\nonumber \\
 & -g_{\circ}\Bigl(\alpha,\frac{-\bigl(R_{\mathrm{x}}-\left\Vert \mathbf{x}_{\mathrm{rx}}\right\Vert \bigr)}{\zeta}\Bigr)\bm{1}_{\left\Vert \mathbf{x}_{\mathrm{rx}}\right\Vert >R_{\mathrm{x}}},\label{EQ9: X(Rx)}
\end{align}
in which $\mathbf{1}_{\mathsf{Q}}=1$ when the statement $\mathsf{Q}$
is true and 0 otherwise. The function $g_{\circ}\left(\alpha,\vartheta\right):\mathbb{R}^{+}\times\mathbb{R}_{0}^{+}\mapsto\mathbb{R}_{0}^{+}$
is defined as:
\begin{equation}
g_{\circ}\left(\alpha,\vartheta\right)\triangleq\intop_{0}^{\vartheta}\frac{\mathrm{d}u}{\left(1+u^{\alpha}\right)}=\vartheta{}_{2}F_{1}\left(1,\frac{1}{\alpha};\bigl(1+\frac{1}{\alpha}\bigr);-\vartheta^{\alpha}\right),\label{EQ10: G-Function}
\end{equation}
in which $_{2}F_{1}$ is Gauss's hypergeometric function. We note
that for certain values of $\alpha>1$, (\ref{EQ10: G-Function})
reverts to a simple form (e.g., $g_{\circ}\left(2,\vartheta\right)=\arctan\left(\vartheta\right)$).

\subsection{Contribution for Vertical Road \textendash{} $\mathcal{P}_{\mathrm{y}}$}

To derive $\mathcal{P}_{\mathrm{y}}$, we notice that the RX and interferers
are accordingly located at $\mathbf{x}_{\mathrm{rx}}=\left[x_{\mathrm{rx}},0\right]^{\mathrm{T}}$
and $\mathbf{x}=\left[0,y\right]^{\mathrm{T}}$. From (\ref{eq:WLOSNLOS})
and (\ref{PYTHM}), we obtain:
\begin{align}
 & \mathcal{P}_{\mathrm{y}}=\exp\biggl(-\int\limits _{-R_{\mathrm{y}}}^{+R_{\mathrm{y}}}\frac{p_{\mathrm{I}}\lambda_{\mathrm{y}}}{1+1/(\beta^{\prime}\ell([0,y]^{\mathrm{T}},\mathbf{x}_{\mathrm{rx}}))}\mathrm{d}y\biggr)\mathrm{.}\label{EQ11: Py}
\end{align}
We show in Appendix \ref{sec:Expression-forPy} that $\mathcal{P}_{\mathrm{y}}=\exp\bigl(-2p_{\mathrm{I}}\lambda_{\mathrm{y}}\zeta\mathcal{Y}(R_{\mathrm{y}})\bigr)$,
where
\begin{align}
 & \mathcal{Y}(R_{\mathrm{y}})=-g_{\circ}\Bigl(\alpha,\frac{\left\Vert \mathbf{x}_{\mathrm{rx}}\right\Vert }{\zeta}\Bigr)\nonumber \\
 & +g_{\circ}\Bigl(\alpha,\frac{\bigl(R_{\mathrm{y}}+\left\Vert \mathbf{x}_{\mathrm{rx}}\right\Vert \bigr)}{\zeta}\Bigr)\bm{1}_{\left\Vert \mathbf{x}_{\mathrm{rx}}\right\Vert \leq\triangle}\nonumber \\
 & +g_{\circ}\Bigl(\alpha,\frac{\bigl(\triangle+\left\Vert \mathbf{x}_{\mathrm{rx}}\right\Vert \bigr)}{\zeta}\Bigr)\bm{1}_{\left\Vert \mathbf{x}_{\mathrm{rx}}\right\Vert >\triangle}\nonumber \\
 & +\frac{1}{\kappa}\biggl(g_{\circ}\Bigl(\alpha,\frac{\kappa R_{\mathrm{y}}}{\zeta}\Bigr)-g_{\circ}\Bigl(\alpha,\frac{\kappa\triangle}{\zeta}\Bigr)\biggr)\bm{1}_{\left\Vert \mathbf{x}_{\mathrm{rx}}\right\Vert >\triangle}\label{eq:YRY}
\end{align}
and $\kappa=\left(A_{\circ}/A_{\circ}^{\prime}\right)^{1/\alpha}\left\Vert \mathbf{x}_{\mathrm{rx}}\right\Vert $.

\section{Analysis and Results}

\begin{table}
\caption{Parameters for Numerical Evaluation\label{TAB: Numerical Parameters}}

\resizebox{0.5\textwidth}{!}{%

\begin{tabular}{ll}
\hline 
\multicolumn{2}{l}{\textbf{System Parameters}}\tabularnewline
\hline 
\hline 
Target Reliability & $\mathcal{P}_{\mathrm{target}}=0.9$\tabularnewline
Transmit Power (dBmW) & $P_{\circ}=20$\tabularnewline
AWGN Floor (dBmW) & $N_{\circ}=-99$\tabularnewline
RX Sensitivity (dB) & $\beta=8$\tabularnewline
\hline 
\hline 
\multicolumn{2}{l}{\textbf{Propagation Parameters}}\tabularnewline
\hline 
\hline 
Operating Frequency (GHz) & $f_{\circ}=5.9$\tabularnewline
Reference Distance (m) & $d_{\circ}=10$\tabularnewline
Break-Point Distance (m) & $\triangle=15$\tabularnewline
Path Loss (PL) Exponent & $\alpha=1.68$\tabularnewline
PL Coefficient for LOS/WLOS (dBm) & $A_{\circ}\!=\!-37.86+10\alpha$\tabularnewline
PL Coefficient for NLOS (dBm), $r\!\in\!\left(0,1\right)$ & $A_{\circ}^{\prime}\!=\!-37.86+7\alpha+10\log_{10}\!\left(r\!\cdot\!\triangle^{\alpha}\right)$\tabularnewline
\hline 
\hline 
\multicolumn{2}{l}{\textbf{TX/RX Geometry}}\tabularnewline
\hline 
\hline 
RX Distance from Intersection (m) & $\left\Vert \mathbf{x}_{\mathrm{rx}}\right\Vert =50$\tabularnewline
Max. TX/RX Manhattan Separation (m) & $D_{\mathrm{max}}=120$\tabularnewline
\hline 
\multicolumn{2}{l}{\textbf{Traffic Parameters of Interferers}}\tabularnewline
\hline 
\hline 
Vehicular Traffic Intensity (\# / m) & $\lambda=0.01$\tabularnewline
Max. Interference Radius (m) & $R_{\mathrm{max}}=1,000$\tabularnewline
\hline 
\end{tabular}

}
\end{table}
\begin{figure}
\begin{centering}
\includegraphics[width=1\columnwidth]{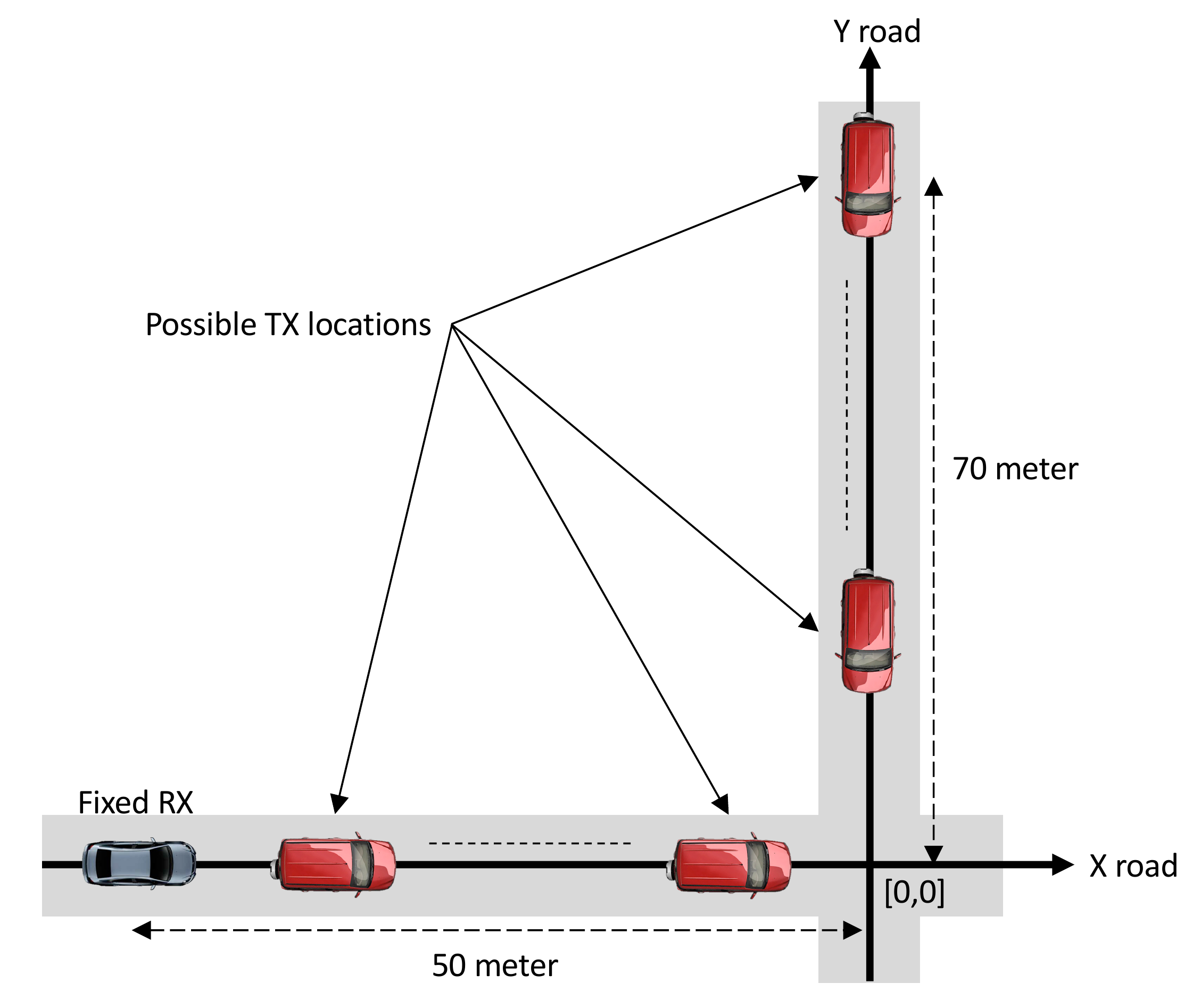}
\par\end{centering}
\caption{TX/RX setup for numerical evaluation with a fixed RX position and
multiple possible TX positions. Interferers are not shown. \label{Fig2: TX/RX Setup(visio)}}
\end{figure}

\subsection{Simulation Setup}

We evaluated the success probability for various scenarios of TX/RX
positions with the parameters shown in Table \ref{TAB: Numerical Parameters}.
We set $\lambda_{\mathrm{x}}=\lambda_{\mathrm{y}}=\lambda=0.01$ and
$R_{\mathrm{x}}=R_{\mathrm{y}}=R\in[\triangle,R_{\mathrm{max}}]$.
Due to the nature of its channel model in (\ref{eq:WLOSNLOS}), we
will determine success probability as a function of the Manhattan
distance (denoted as $\left\Vert \cdot\right\Vert _{1}$ for the $l_{1}$-norm).
In particular, we consider a fixed RX on the horizontal road at $\mathbf{x}_{\mathrm{rx}}=[-50\text{\quotesinglbase0}]^{\mathrm{T}}$
and a TX that can be in different positions up to a Manhattan distance
of $D_{\mathrm{max}}=120\,\mathrm{m}$ on the vertical road (see Fig.~\ref{Fig2: TX/RX Setup(visio)}).

In terms of design, we will aim to achieve a target success probability
$\mathcal{P}_{\mathrm{target}}\in(0,1)$, generally close to 1, over
a certain area. In other words, we want 
\begin{equation}
\mathcal{P}_{\mathrm{noint}}\mathcal{P}_{\mathrm{x}}\mathcal{P}_{\mathrm{y}}\ge\mathcal{P}_{\mathrm{target}},\label{eq:designtarget}
\end{equation}
for all $\mathbf{x}_{\mathrm{rx}}$ and $\mathbf{x}_{\mathrm{tx}}$
under consideration. As design parameters, we will consider the Aloha
transmit probability $p_{\mathrm{I}}$ and the interference range
$R$. 

\subsection{Design: Aloha Transmit Probability vs Interference Range}

\begin{figure}
\begin{centering}
\includegraphics[width=1\columnwidth]{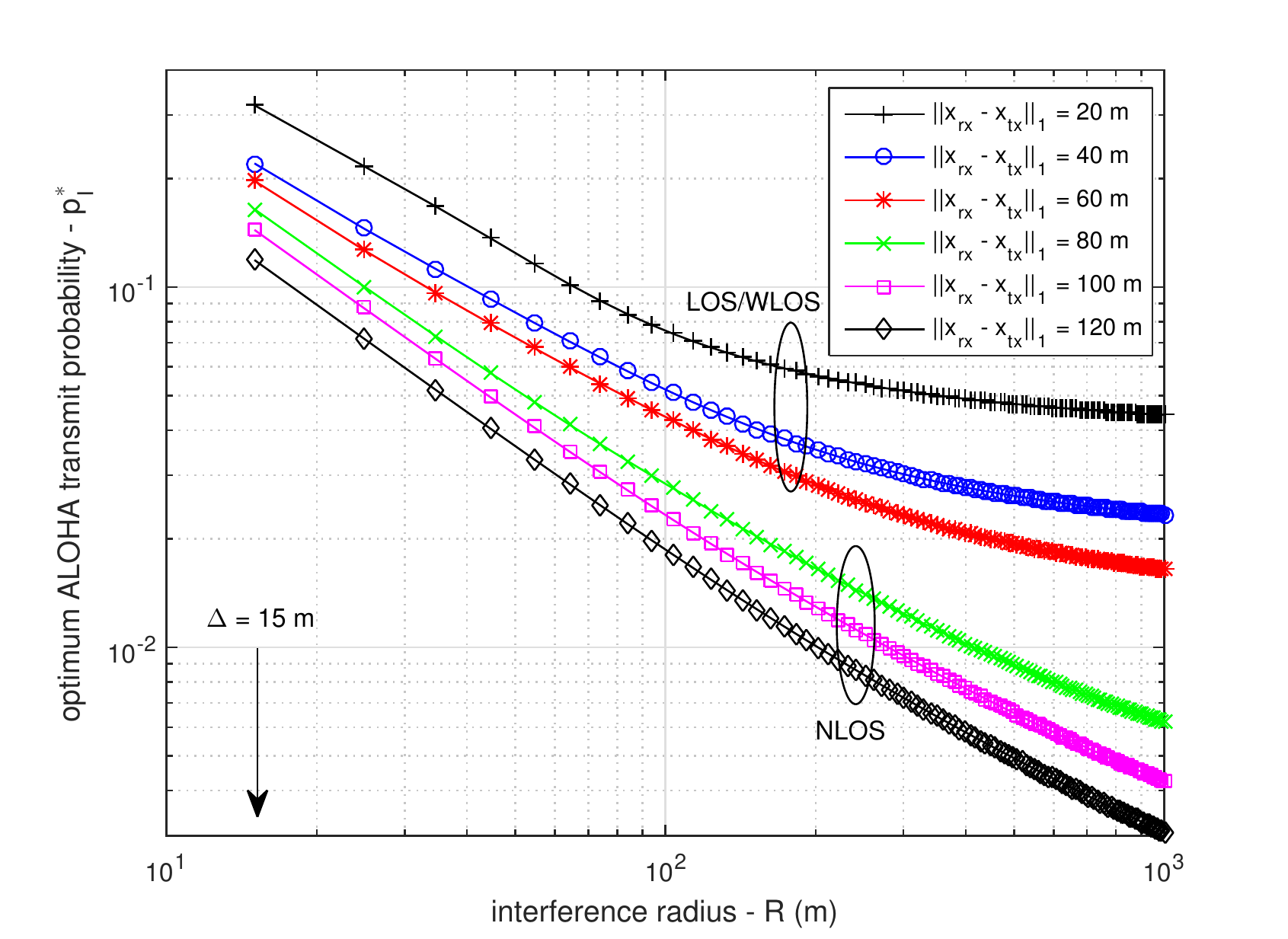}
\par\end{centering}
\caption{Optimal Aloha transmit probability as a function of interference radius
over different values of TX/RX separation.\label{Fig3: Pi*vs.R(matlab)}}
\end{figure}

We will first determine an optimal Aloha transmit probability as a
function of the interference range $R$, for a given target performance
requirement, $\mathcal{P}_{\mathrm{target}}$. Solving (\ref{eq:designtarget})
for $p_{\mathrm{I}}$ yields 
\begin{align}
 & p_{\mathrm{I}}^{\ast}\left(R\right)\!=\!\frac{-\beta N_{\circ}/(P_{\circ}\ell(\mathbf{x}_{\mathrm{tx}},\mathbf{x}_{\mathrm{rx}}))-\ln(\mathcal{P}_{\mathrm{target}})}{\zeta\lambda(\mathcal{X}\left(R\right)+2\mathcal{Y}\left(R\right))}.\label{EQ14: Optimal Transmit Probability-1}
\end{align}
This relationship is shown in Fig.~\ref{Fig3: Pi*vs.R(matlab)} for
different values of $\mathbf{x}_{\mathrm{tx}}$ (and thus of $\Vert\mathbf{x}_{\mathrm{rx}}-\mathbf{x}_{\mathrm{tx}}\Vert_{1}$).
We observe that $p_{\mathrm{I}}^{\ast}\left(R\right)$ is monotonically
decreasing in $R$, since a larger region of possible transmitters
requires a reduction in $p_{\mathrm{I}}$ in order to meet the target
performance. This relationship also shows that as the RX remains
fixed and the TX moves across: (i) LOS: $\Vert\mathbf{x}_{\mathrm{rx}}-\mathbf{x}_{\mathrm{tx}}\Vert_{1}\in(0,\Vert\mathbf{x}_{\mathrm{rx}}\Vert]$,
(ii) WLOS: $\Vert\mathbf{x}_{\mathrm{rx}}-\mathbf{x}_{\mathrm{tx}}\Vert_{1}\in(\Vert\mathbf{x}_{\mathrm{rx}}\Vert,\Vert\mathbf{x}_{\mathrm{rx}}\Vert+\Delta]$;
and (iii) NLOS: $\Vert\mathbf{x}_{\mathrm{rx}}-\mathbf{x}_{\mathrm{tx}}\Vert_{1}\in(\Vert\mathbf{x}_{\mathrm{rx}}\Vert+\Delta,D_{\mathrm{max}}]$,
a better channel environment (such as LOS and WLOS) can tolerate more
active interfering nodes (i.e., a larger $p_{\mathrm{I}}^{\ast}$)
than in severe NLOS situations.

\subsubsection*{Remark}

The expression (\ref{EQ14: Optimal Transmit Probability-1}) is only
valid when $p_{\mathrm{I}}^{\ast}\left(R\right)\ge0$. It is readily
verified that, since the denominator of (\ref{EQ14: Optimal Transmit Probability-1})
is positive, this is equivalent to the natural condition $\mathcal{P}_{\mathrm{noint}}\ge\mathcal{P}_{\mathrm{target}}$,
i.e., the target reliability in the presence of interference can not
exceed the success probability of the wanted TX/RX communication link
under no interference. For the value $D_{\mathrm{max}}$ of 120 m
in our scenario (see Fig.~\ref{Fig2: TX/RX Setup(visio)}), $\mathcal{P}_{\mathrm{noint}}$
turns out to be 0.966, hence $\mathcal{P}_{\mathrm{target}}=0.9$
is a feasible value for all cases under consideration. 

\subsection{Analysis: Sensitivity to TX/RX Separation}

\begin{figure*}[t]
\begin{centering}
\subfloat[$p_{\mathrm{I}}^{\ast}\left(R\right)$ designed for $\left\Vert \mathbf{x}_{\mathrm{rx}}\!-\!\tilde{\mathbf{x}}_{\mathrm{tx}}\right\Vert _{1}\!\leq\!20$\label{Fig4a: up to 20m(matlab)}]{\begin{centering}
\includegraphics[width=0.27\paperwidth]{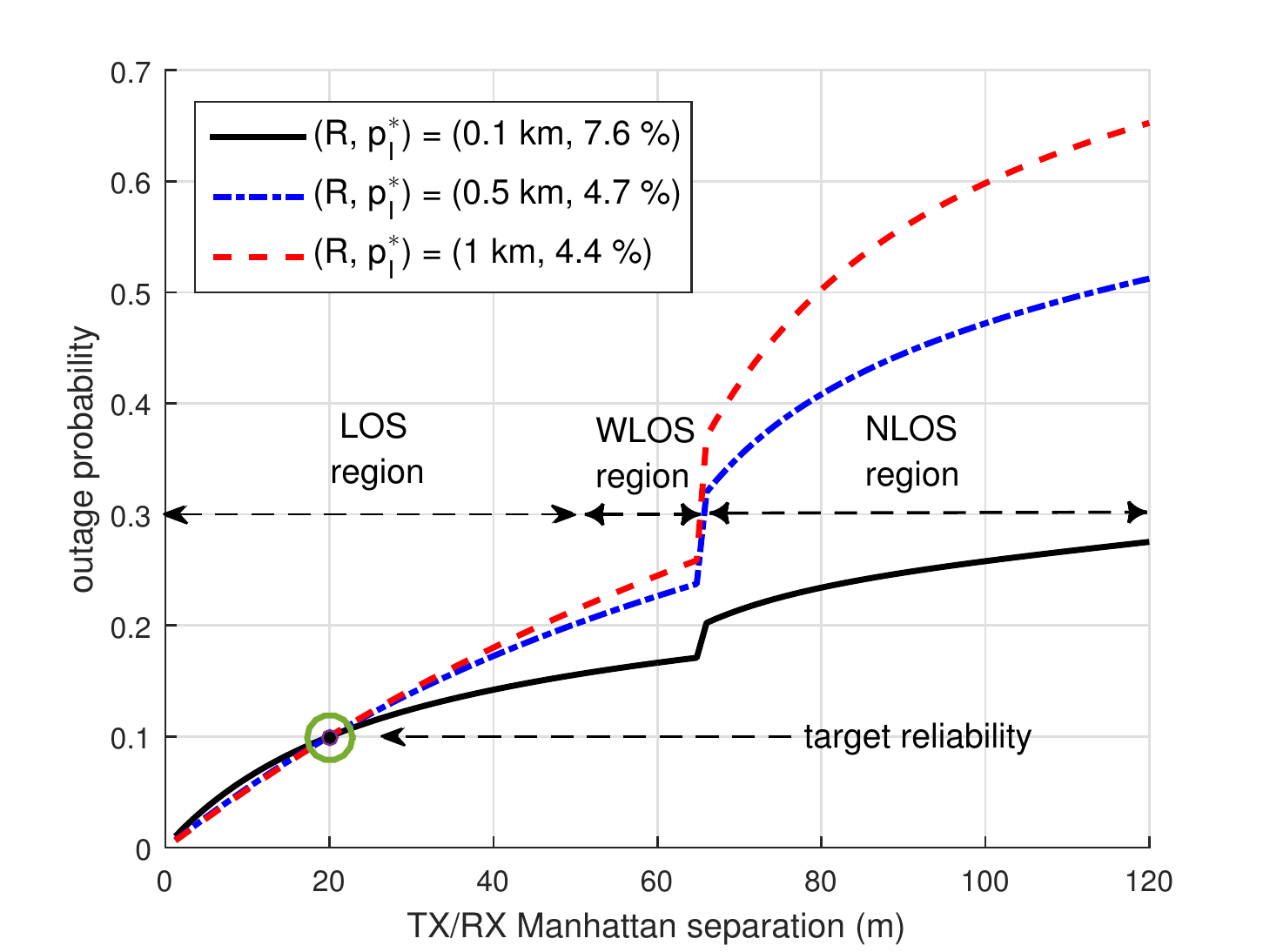}
\par\end{centering}
} \subfloat[$p_{\mathrm{I}}^{\ast}\left(R\right)$ designed for $\left\Vert \mathbf{x}_{\mathrm{rx}}\!-\!\tilde{\mathbf{x}}_{\mathrm{tx}}\right\Vert _{1}\!\leq\!40$\label{Fig4b: up to 40m(matlab)}]{\begin{centering}
\includegraphics[width=0.27\paperwidth]{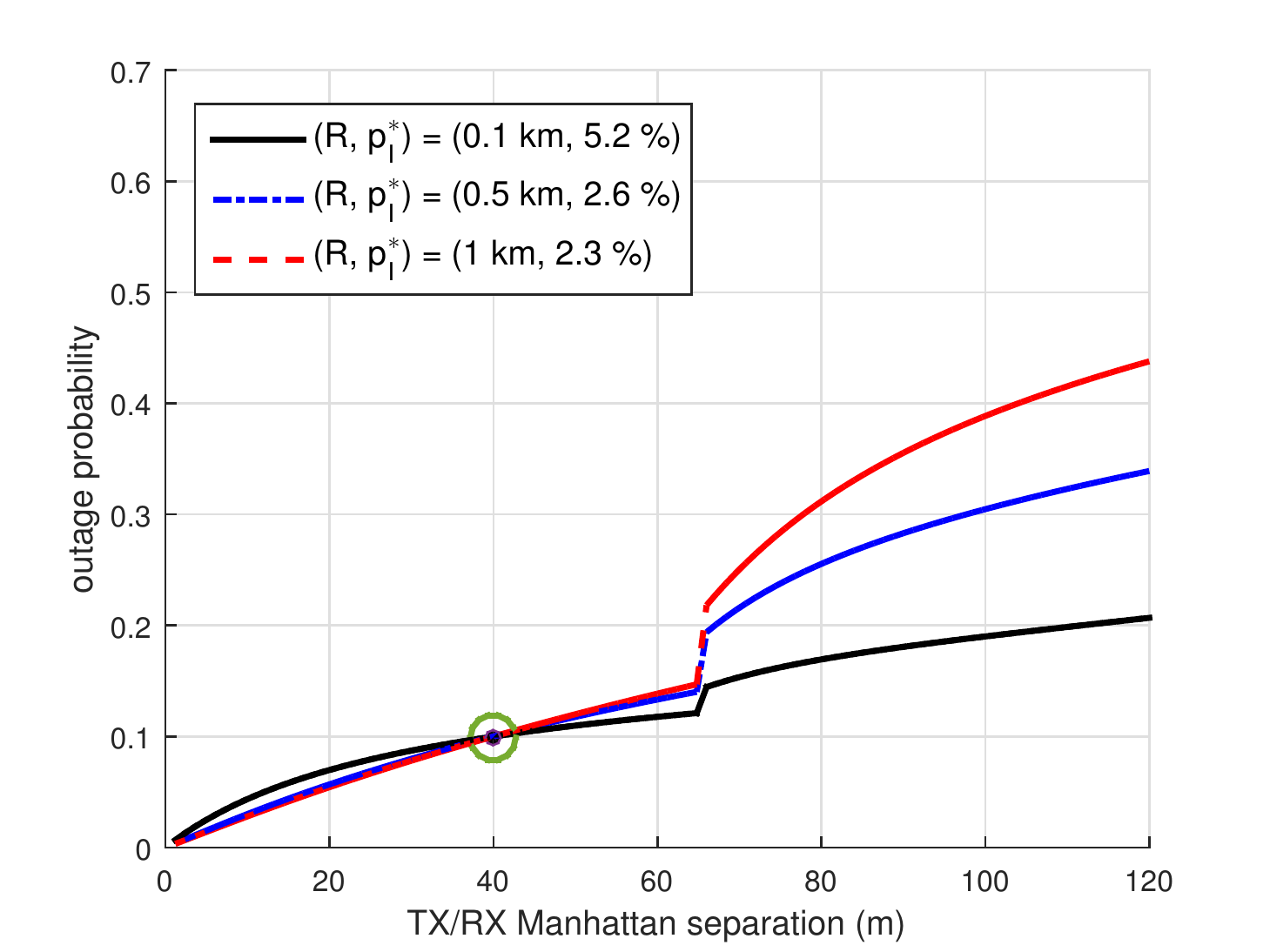}
\par\end{centering}
} \subfloat[$p_{\mathrm{I}}^{\ast}\left(R\right)$ designed for $\left\Vert \mathbf{x}_{\mathrm{rx}}\!-\!\tilde{\mathbf{x}}_{\mathrm{tx}}\right\Vert _{1}\!\leq\!60$\label{Fig4c: up to 60m(matlab)}]{\begin{centering}
\includegraphics[width=0.27\paperwidth]{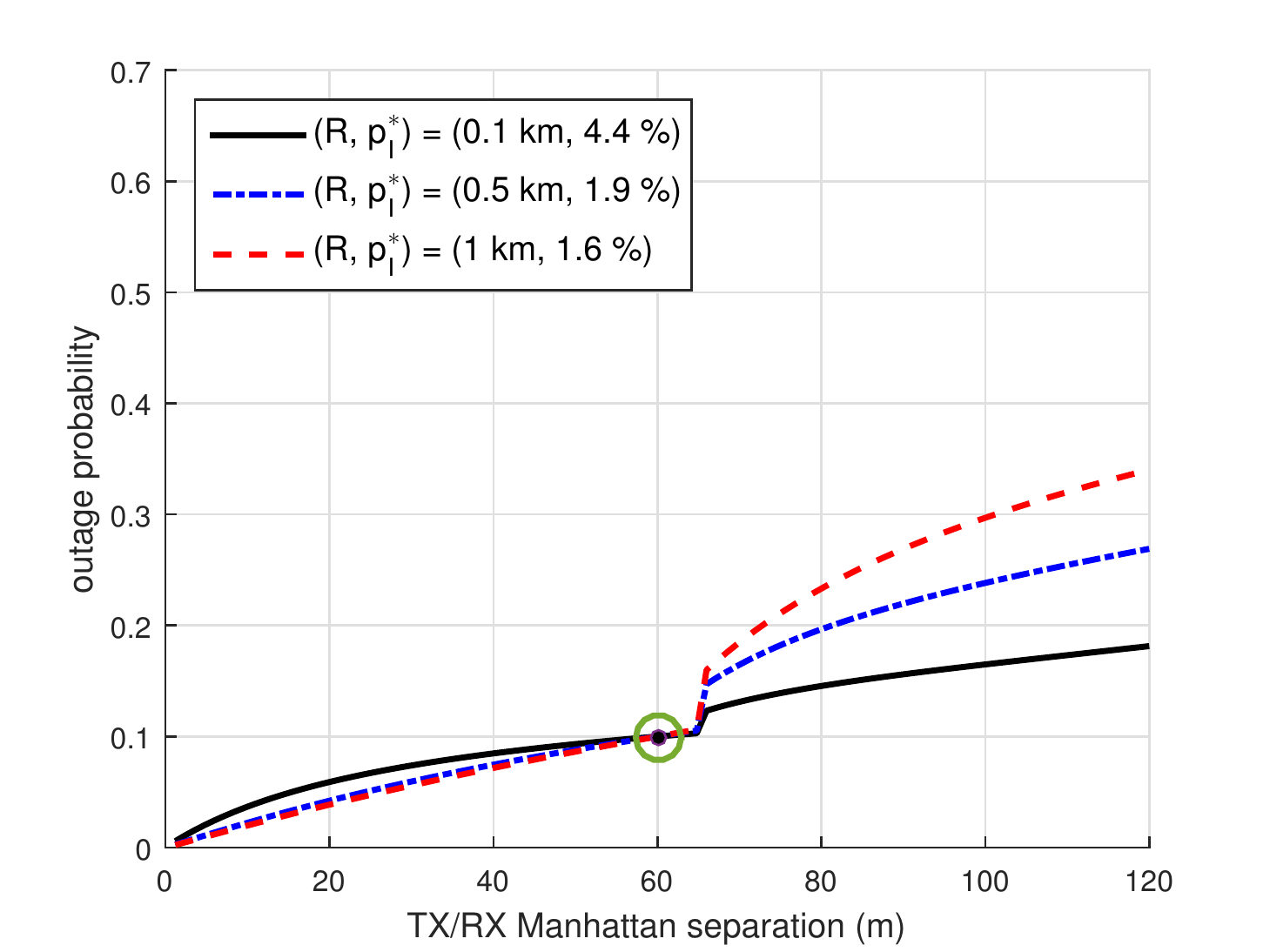}
\par\end{centering}
}
\par\end{centering}
\begin{centering}
\subfloat[$p_{\mathrm{I}}^{\ast}\left(R\right)$ designed for $\left\Vert \mathbf{x}_{\mathrm{rx}}\!-\!\tilde{\mathbf{x}}_{\mathrm{tx}}\right\Vert _{1}\!\leq\!80$\label{Fig4d: up to 80m(matlab)}]{\begin{centering}
\includegraphics[width=0.27\paperwidth]{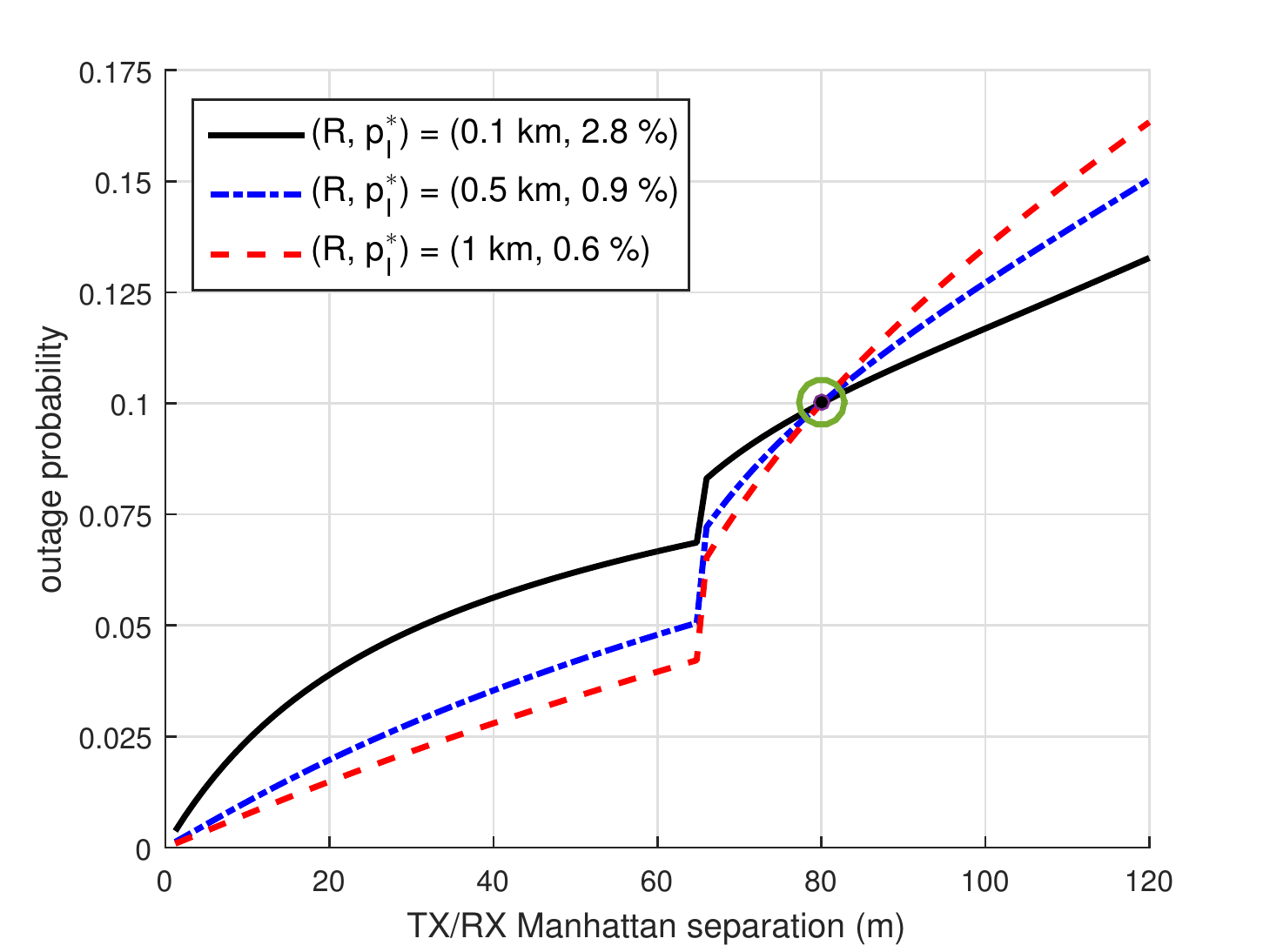}
\par\end{centering}
}\subfloat[achieving $\mathcal{P}_{\mathrm{target}}$ for: $\left\Vert \mathbf{x}_{\mathrm{rx}}\!-\!\tilde{\mathbf{x}}_{\mathrm{tx}}\right\Vert _{1}\!\leq\!100$\label{Fig4e: up to 100m(matlab)}]{\begin{centering}
\includegraphics[width=0.27\paperwidth]{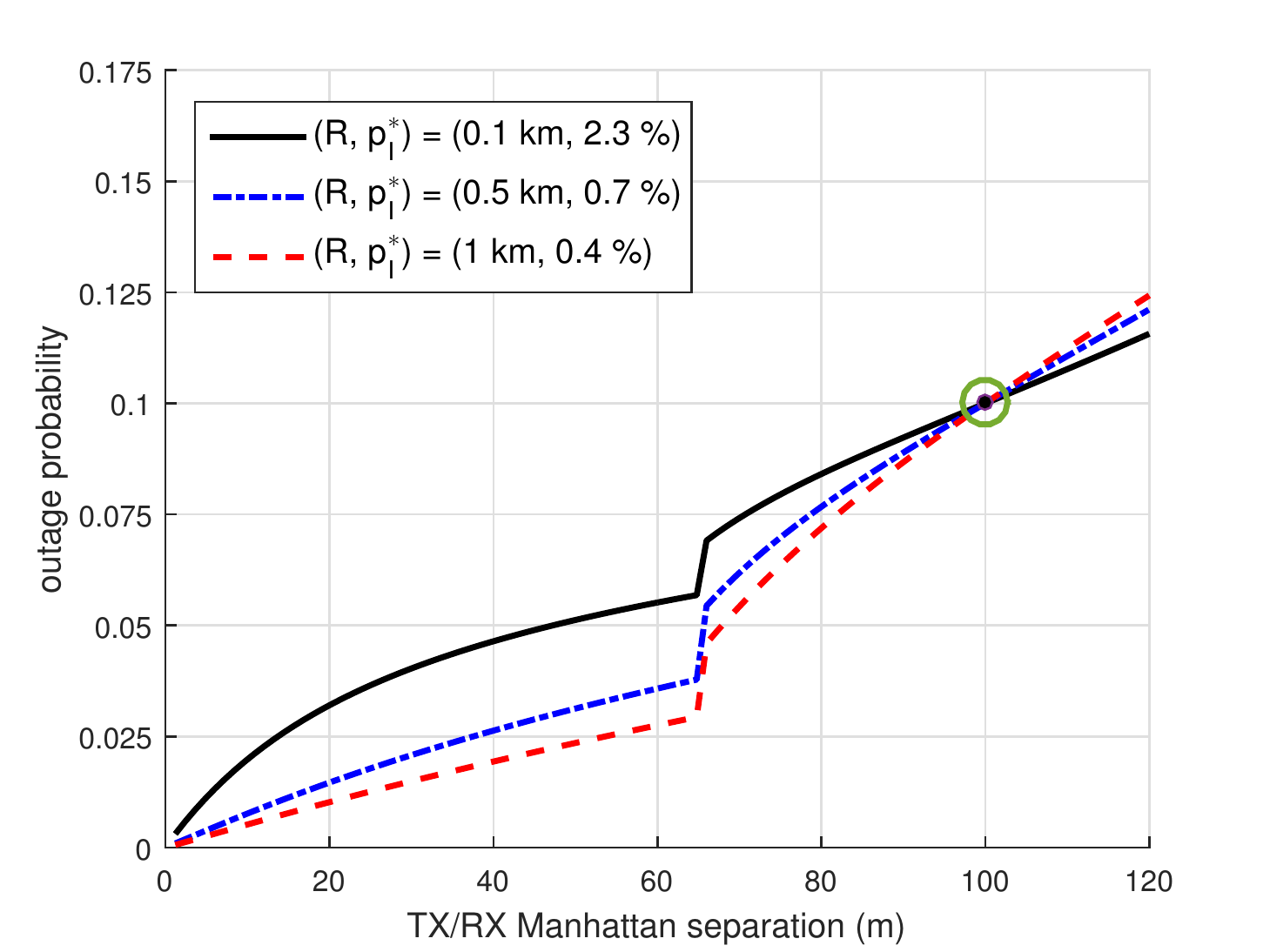}
\par\end{centering}
}\subfloat[$p_{\mathrm{I}}^{\ast}\left(R\right)$ designed for $\left\Vert \mathbf{x}_{\mathrm{rx}}\!-\!\tilde{\mathbf{x}}_{\mathrm{tx}}\right\Vert _{1}\!\leq\!120$\label{Fig4f: up to 120m(matlab)}]{\begin{centering}
\includegraphics[width=0.27\paperwidth]{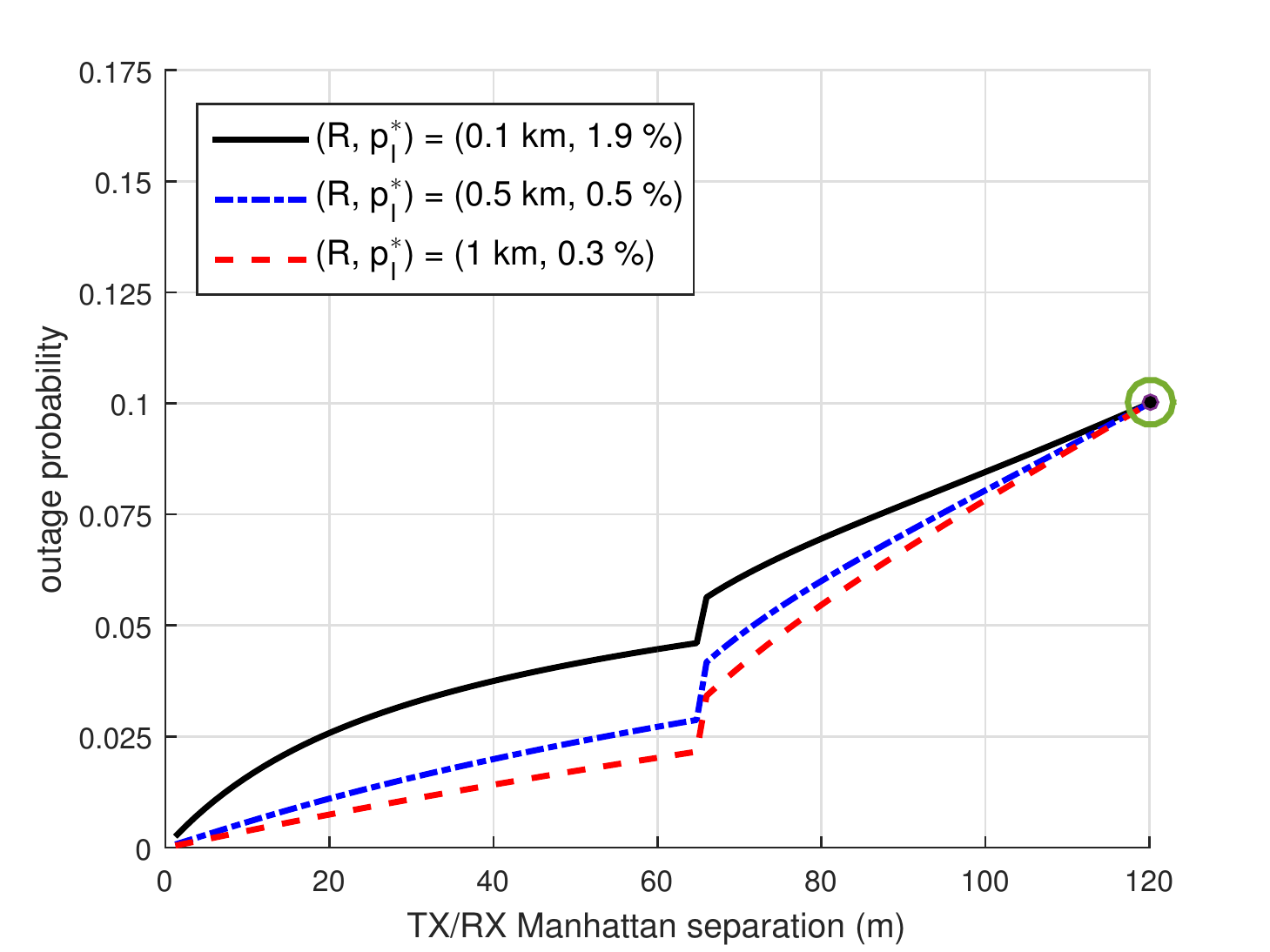}
\par\end{centering}
}
\par\end{centering}
\centering{}\caption{Sensitivity of the outage probability to the TX/RX separation for
different design choices. \label{Fig4: Pout vs. TX/RX Separation (matlab)}}
\end{figure*}

The design from (\ref{EQ14: Optimal Transmit Probability-1}) considers
a given $R$ and a certain $\mathbf{x}_{\mathrm{tx}}$ and $\mathbf{x}_{\mathrm{rx}}$.
In this section, we will evaluate the sensitivity of the success probability
when the TX and RX are in different locations. In particular, we determine
$p_{\mathrm{I}}^{\ast}\left(R\right)$ for $R\in\{100,500,1000\}$,
$\mathbf{x}_{\mathrm{rx}}=[-50,0]^{\mathrm{T}}$, and $\Vert\mathbf{x}_{\mathrm{rx}}-\tilde{\mathbf{x}}_{\mathrm{tx}}\Vert_{1}\in\{20,40,60,80,100,120\}$,
corresponding to $\tilde{\mathbf{x}}_{\mathrm{tx}}\in\{[-30,0]^{\mathrm{T}},[-10,0]^{\mathrm{T}},[0,10]^{\mathrm{T}},[0,30]^{\mathrm{T}},[0,50]^{\mathrm{T}},[0,70]^{\mathrm{T}}\}$.
For these designs, we can then compute $\mathcal{P}_{\mathrm{c}}\left(\beta,\mathbf{x}_{\text{\ensuremath{\mathrm{tx}}}},\mathbf{x}_{\mathrm{rx}}\right)$
for any $\mathbf{x}_{\mathrm{tx}}$ under consideration. For visualization
purposes, we plot the \emph{outage probability}, defined as $1-\mathcal{P}_{\mathrm{c}}\left(\beta,\mathbf{x}_{\text{\ensuremath{\mathrm{tx}}}},\mathbf{x}_{\mathrm{rx}}\right)$,
as a function of TX/RX Manhattan separation in Fig.~\ref{Fig4: Pout vs. TX/RX Separation (matlab)}. 

To understand the figure, take for example Fig.~\ref{Fig4a: up to 20m(matlab)},
where $p_{\mathrm{I}}^{\ast}\left(R\right)$ was determined for $R\in\{100,500,1000\}$,
$\tilde{\mathbf{x}}_{\mathrm{tx}}=[-30,0]^{\mathrm{T}}$ and $\mathbf{x}_{\mathrm{rx}}\!=\!\left[-50,0\right]^{\mathrm{T}}$.
For this $p_{\mathrm{I}}^{\ast}\left(R\right)$, we show the outage
probability as a function of $\Vert\mathbf{x}_{\mathrm{rx}}-\mathbf{x}_{\mathrm{tx}}\Vert_{1}$,
for our scenario, depicted in Fig.~\ref{Fig2: TX/RX Setup(visio)}. 

We first note that $\Vert\mathbf{x}_{\mathrm{rx}}-\mathbf{x}_{\mathrm{tx}}\Vert_{1}=\Vert\mathbf{x}_{\mathrm{rx}}\!-\!\tilde{\mathbf{x}}_{\mathrm{tx}}\Vert_{1}$,
the packet reliability of 0.9 (shown with a green circle mark, for
a corresponding outage of 0.1) is achieved. When $\Vert\mathbf{x}_{\mathrm{rx}}-\mathbf{x}_{\mathrm{tx}}\Vert_{1}<\Vert\mathbf{x}_{\mathrm{rx}}\!-\!\tilde{\mathbf{x}}_{\mathrm{tx}}\Vert_{1}$,
the outage reduces, while for $\Vert\mathbf{x}_{\mathrm{rx}}-\mathbf{x}_{\mathrm{tx}}\Vert_{1}>\Vert\mathbf{x}_{\mathrm{rx}}\!-\!\tilde{\mathbf{x}}_{\mathrm{tx}}\Vert_{1}$,
the outage increases. For each of the subfigures, the three curves
(corresponding to different values of $R$), we observe a distinctive
format, consistent with the uniqueness of the urban intersection path
loss models. Due to the non-continuous nature of model (\ref{eq:WLOSNLOS}),
the outage curves show a discontinuity when $\mathbf{x}_{\mathrm{tx}}$
transitions from WLOS to NLOS (this happens when $\left\Vert \mathbf{x}_{\mathrm{rx}}-\mathbf{x}_{\mathrm{tx}}\right\Vert _{1}=\left\Vert \mathbf{x}_{\mathrm{rx}}\right\Vert +\triangle$,
which in our case corresponds to a separation of 65 m).

Secondly, we note that the smallest interference region (i.e., $R=100\,\mathrm{m}$)
corresponds to the largest transmit probability. This smallest interference
region leads to the \emph{largest} outages for $\left\Vert \mathbf{x}_{\mathrm{rx}}\!-\!\mathbf{x}_{\mathrm{tx}}\right\Vert _{1}<\Vert\mathbf{x}_{\mathrm{rx}}\!-\!\tilde{\mathbf{x}}_{\mathrm{tx}}\Vert_{1}$,
though never surpassing 0.1. This is due to the larger possibility
of active transmitters in close proximity to the RX. On the other
hand, the smallest interference region leads to the \emph{smallest}
outages for $\left\Vert \mathbf{x}_{\mathrm{rx}}\!-\!\mathbf{x}_{\mathrm{tx}}\right\Vert _{1}>\Vert\mathbf{x}_{\mathrm{rx}}\!-\!\tilde{\mathbf{x}}_{\mathrm{tx}}\Vert_{1}$.
This is because the outage is dominated by the aggregate interference,
rather than the interferers close to the RX. Hence, the larger interference
region, which has more interferers, has the largest outages. In other
words, a small interference region allows for a high density of active
transmitters $\lambda p_{\mathrm{I}}$, while leading to relatively
graceful degradation outside the interference region. 

Finally, we see that as $\tilde{\mathbf{x}}_{\mathrm{tx}}$ is varied
in the different plots in Fig.~\ref{Fig4: Pout vs. TX/RX Separation (matlab)},
the optimal $p_{\mathrm{I}}^{\ast}\left(R\right)$ varies significantly.
In particular, comparing the values of $p_{\mathrm{I}}^{\ast}\left(R\right)$
for $\tilde{\mathbf{x}}_{\mathrm{tx}}=[-30,0]^{\mathrm{T}}$ (Fig.~\ref{Fig4a: up to 20m(matlab)})
with $\tilde{\mathbf{x}}_{\mathrm{tx}}=[0,70]^{\mathrm{T}}$ (Fig.~\ref{Fig4f: up to 120m(matlab)}),
the Aloha transmit probabilities are reduced by a factor of 4 for
$R=100$ and a factor of 15 for $R=1000$. Hence, larger transmission
ranges come at a severe cost of reduced density of active transmitters.
In summary, our analysis indicates that when a system is designed
for a certain maximum communication range (e.g., a Manhattan distance
of 100 m, see Fig.~\ref{Fig4e: up to 100m(matlab)}), it is recommended
to set $R$ as low as possible (in this case $R=50$ m is recommended),
as this leads to the highest density of active transmitters and and
a graceful performance degradation outside the interference region. 

\section{Conclusion}

V2V communication is critical for future intelligent transportation
systems. A key performance metric is the probability of successful
packet delivery in the presence of interference. In this paper, we
analytically characterized the success probability for urban intersections
based on specialized path loss models. It turns out that these path
loss models are amenable for mathematical analysis and lead to exact
closed-form expressions for different path loss exponents and finite
interference regions. As shown in the paper, the derived expressions
can aid in the communication system design task, complementing time-consuming
simulations and experiments. In particular, we found that from a system
perspective, it is beneficial to limit interference to a small spatial
region, while allowing more simultaneous transmitters. 

\appendices{}

\section{Expression for $\mathcal{P}_{\mathrm{x}}$\label{sec:Expression-forPx}}

\subsubsection*{Case I \textendash{} RX is Inside $B_{\mathrm{x}}$ (i.e., $\left\Vert \mathbf{x}_{\mathrm{rx}}\right\Vert \leq R_{\mathrm{x}}$)}

Due to $\left|x_{\mathrm{rx}}-x\right|$, the integral (\ref{PXTHM})
must be split in two parts, namely from $x\in[-R_{\mathrm{x}},x_{\mathrm{rx}}]$
(for which $\vert x_{\mathrm{rx}}-x\vert=x_{\mathrm{rx}}-x$) and
from $x\in[x_{\mathrm{rx}},+R_{\mathrm{x}}]$ (for which $\vert x_{\mathrm{rx}}-x\vert=x-x_{\mathrm{rx}}$).
If we let $u=(x_{\mathrm{rx}}-x)/\zeta$ for the first part, and
$v=(x-x_{\mathrm{rx}})/\zeta$ for the second, (\ref{PXTHM}) becomes:
\begin{align}
 & \mathcal{P}_{\mathrm{x}}=\label{EQ-App-A: 2}\\
 & \exp\biggl(-p_{\mathrm{I}}\lambda_{\mathrm{x}}\zeta\biggl\{ g_{\circ}\Bigl(\alpha,\frac{\left(R_{\mathrm{x}}+x_{\mathrm{rx}}\right)}{\zeta}\Bigr)+g_{\circ}\Bigl(\alpha,\frac{\left(R_{\mathrm{x}}-x_{\mathrm{rx}}\right)}{\zeta}\Bigr)\biggr\}\biggr).\nonumber 
\end{align}
Meanwhile, we should underscore that due to the symmetry in (\ref{EQ-App-A: 2}),
it is possible to replace $x_{\mathrm{rx}}$ by $\left\Vert \mathbf{x}_{\mathrm{rx}}\right\Vert $,
while still remaining compatible when $x_{\mathrm{rx}}<0$.\\

\subsubsection*{Case II \textendash{} RX is Outside $B_{\mathrm{x}}$ (i.e., $\left\Vert \mathbf{x}_{\mathrm{rx}}\right\Vert >R_{\mathrm{x}}$)}

The RX must be outside the region of H-PPP interferers on road-$x$;
therefore, we may consider $x_{\mathrm{rx}}<-R_{\mathrm{x}}$ or $x_{\mathrm{rx}}>R_{\mathrm{x}}$.
Due to symmetry, the final result will be identical. Considering the
RX positioned on the negative axis, we replace $\left|x_{\mathrm{rx}}-x\right|$
by $\left(x-x_{\mathrm{rx}}\right)$ in (\ref{EQ8: Px}), while taking
the integration over $\left|x\right|\leq R_{\mathrm{x}}$; also, realizing
that $-x_{\mathrm{rx}}=\left\Vert \mathbf{x}_{\mathrm{rx}}\right\Vert $,
we get:
\begin{align}
 & \mathcal{P}_{\mathrm{x}}=\exp\biggl(-\intop_{-R_{\mathrm{x}}}^{R_{\mathrm{x}}}\frac{p_{\mathrm{I}}\lambda_{\mathrm{x}}}{1+\bigl(\bigl(x+\Vert\mathbf{x}_{\mathrm{rx}}\vert\bigr)/\zeta\bigr)^{\alpha}}\mathrm{d}x\biggr)\mathrm{.}\label{EQ-App-A: 3}
\end{align}
If we let $u=\left(x+\left\Vert \mathbf{x}_{\mathrm{rx}}\right\Vert \right)/\zeta$,
the expression in (\ref{EQ-App-A: 3}) will then equal to:
\begin{align}
 & \mathcal{P}_{\mathrm{x}}=\exp\biggl(-p_{\mathrm{I}}\lambda_{\mathrm{x}}\zeta\!\!\!\!\!\!\!\!\!\!\!\intop_{\left(\left\Vert \mathbf{x}_{\mathrm{rx}}\right\Vert -R_{\mathrm{x}}\right)/\zeta}^{\left(\left\Vert \mathbf{x}_{\mathrm{rx}}\right\Vert +R_{\mathrm{x}}\right)/\zeta}\!\!\!\!\!\frac{\mathrm{d}u}{\bigl(1+u^{\alpha}\bigr)}\biggr)=\label{EQ-App-A: 4}\\
 & \!\!\exp\biggl(\!\!-p_{\mathrm{I}}\lambda_{\mathrm{x}}\zeta\biggl\{\!g_{\circ}\!\Bigl(\!\alpha,\!\frac{\left(\left\Vert \mathbf{x}_{\mathrm{rx}}\right\Vert \!+\!R_{\mathrm{x}}\right)}{\zeta}\!\Bigr)\!-\!g_{\circ}\!\Bigl(\!\alpha,\!\frac{\left(\left\Vert \mathbf{x}_{\mathrm{rx}}\right\Vert \!-\!R_{\mathrm{x}}\right)}{\zeta}\!\Bigr)\!\!\biggr\}\!\!\biggr)\mathrm{.}\nonumber 
\end{align}

\section{Expression for $\mathcal{P}_{\mathrm{y}}$\label{sec:Expression-forPy}}

\subsubsection*{Case I \textendash{} RX is Near the Intersection (i.e., $\left\Vert \mathbf{x}_{\mathrm{rx}}\right\Vert \leq\triangle$)}

When the RX is close to the intersection, the WLOS Manhattan model
within (\ref{eq:WLOSNLOS}) is relevant:
\begin{align}
 & \mathcal{P}_{\mathrm{y}}=\exp\biggl(-\int_{B_{\mathrm{y}}}\!\!\frac{p_{\mathrm{I}}\lambda_{\mathrm{y}}\ \mathrm{d}y}{\bigl(1+\bigl(\!\bigl(\left|y\right|\!+\!\left\Vert \mathbf{x}_{\mathrm{rx}}\right\Vert \bigr)/\zeta\bigr)^{\alpha}\bigr)}\biggr)\label{EQ-App-B: 1}
\end{align}
where $\zeta=\left(A_{\circ}\beta^{\prime}\right)^{1/\alpha}$. If
we perform a change of variable to (\ref{EQ-App-B: 1}) with $u=\bigl(\left|y\right|+\left\Vert \mathbf{x}_{\mathrm{rx}}\right\Vert \bigr)/\zeta$,
we obtain:
\begin{align}
 & \mathcal{P}_{\mathrm{y}}=\label{EQ-App-B: 2}\\
 & \exp\biggl(\!-2p_{\mathrm{I}}\lambda_{\mathrm{y}}\zeta\biggl\{\!g_{\circ}\Bigl(\!\alpha,\frac{\bigl(R_{\mathrm{y}}+\left\Vert \mathbf{x}_{\mathrm{rx}}\right\Vert \bigr)}{\zeta}\!\Bigr)-g_{\circ}\Bigl(\!\alpha,\frac{\left\Vert \mathbf{x}_{\mathrm{rx}}\right\Vert }{\zeta}\!\Bigr)\!\biggr\}\!\biggr).\nonumber 
\end{align}
\\

\subsubsection*{Case II \textendash{} RX is Away from the Intersection (i.e., $\left\Vert \mathbf{x}_{\mathrm{rx}}\right\Vert >\triangle$)}

In this case, the WLOS Manhattan model within (\ref{eq:WLOSNLOS})
is relevant only when $\left\Vert \mathbf{x}\right\Vert \leq\triangle$,
while the NLOS should be used when $\triangle\!<\!\left\Vert \mathbf{x}\right\Vert \!\leq\!R_{\mathrm{y}}$.
Applying these models into (\ref{EQ11: Py}), we get
\begin{align}
 & \negthinspace\negthinspace\mathcal{P}_{\mathrm{y}}\!=\!\exp\Biggl(\!-2p_{\mathrm{I}}\lambda_{\mathrm{y}}\Biggl\{\intop_{0}^{\triangle}\!\frac{\mathrm{d}y}{\left(1+\left(\left(y+\left\Vert \mathbf{x}_{\mathrm{rx}}\right\Vert \right)/\zeta\right)^{\alpha}\right)}\nonumber \\
 & \ \ \ \ \ \ \ \ \ \ \ \ \ \ \ \ \ \ \ \ \ \ \ +\ \intop_{\triangle}^{R_{y}}\frac{\mathrm{d}y}{\left(1+\left(y\cdot\left\Vert \mathbf{x}_{\mathrm{rx}}\right\Vert /\zeta^{\prime}\right)^{\alpha}\right)}\Biggr\}\Biggr)\label{EQ-App-B: 3}
\end{align}
where $\zeta\negthinspace=\negthinspace\left(A_{\circ}\beta^{\prime}\right)^{1/\alpha}$
and $\zeta^{\prime}\negthinspace=\negthinspace\left(A_{\circ}^{\prime}\beta^{\prime}\right)^{1/\alpha}\negthinspace=\negthinspace\zeta\left(A_{\circ}^{\prime}/A_{\circ}\right)^{1/\alpha}$.
If we let $u=\left(y+\left\Vert \mathbf{x}_{\mathrm{rx}}\right\Vert \right)/\zeta$
for the first integration in (\ref{EQ-App-B: 3}), and $v=y\left\Vert \mathbf{x}_{\mathrm{rx}}\right\Vert /\zeta^{\prime}$
for the second, we get:
\begin{align}
 & \mathcal{P}_{\mathrm{y}}=\exp\biggl(-2p_{\mathrm{I}}\lambda_{\mathrm{y}}\zeta\biggl\{ g_{\circ}\Bigl(\alpha,\frac{\bigl(\triangle+\left\Vert \mathbf{x}_{\mathrm{rx}}\right\Vert \bigr)}{\zeta}\Bigr)-\nonumber \\
 & g_{\circ}\Bigl(\alpha,\frac{\left\Vert \mathbf{x}_{\mathrm{rx}}\right\Vert }{\zeta}\Bigr)\!+\!\frac{1}{\kappa}\biggl(g_{\circ}\Bigl(\alpha,\frac{\kappa R_{y}}{\zeta}\Bigr)-g_{\circ}\Bigl(\alpha,\frac{\kappa\triangle}{\zeta}\Bigr)\biggr)\!\biggr\}\!\biggr)\label{EQ-App-B: 4}
\end{align}
where $\kappa\!=\!\left(A_{\circ}/A_{\circ}^{\prime}\right)^{1/\alpha}\left\Vert \mathbf{x}_{\mathrm{rx}}\right\Vert $
and $g_{\circ}\left(\alpha,\vartheta\right)$ is defined in (\ref{EQ10: G-Function}).

\section*{Acknowledgment}

This research work is supported, in part, by the European Commission
under the Marie Sk\l odowska-Curie Individual Fellowship (H2020-MSCA-IF-2014),
Grant No. 659933 (MARSS-5G); the EU-H2020 project HIGHTS (High Precision
Positioning for Cooperative ITS Applications), Grant No. MG-3.5a-2014-636537;
and VINNOVA under the program \textquotedblleft Nationell Metrologi
vid SP Sveriges Tekniska Forskningsinstitut\textquotedblright , \textquotedblleft COPPLAR
CampusShuttle Cooperative Perception and Planning Platform\textquotedblright ,
funded under Strategic Vehicle Research and Innovation, Grant No.
2015-04849.

\bibliographystyle{IEEEtran}
\bibliography{refGC16}

\end{document}